\documentclass[showpacs,aps,amssymb,floatfix,prd,preprintnumbers]{revtex4-2}
\usepackage{epstopdf}
\usepackage{capt-of}
\usepackage{graphicx}  % Include figure files
\usepackage{dcolumn}   % Align table columns on decimal point
\usepackage{bm}% bold math
\usepackage{amsmath}
\usepackage[font=scriptsize]{caption}
\usepackage[colorlinks]{hyperref}
\begin{document}
\input epsf.tex
%%%%%%%%%%%%
%%%%%%%%%%%%
\title{\bf Accelerating Universe with binary mixture of bulk viscous fluid and dark energy}

\author{Nishant Singla}
\email{nishantsinglag@gmail.com} \affiliation{Department of Physics, Suresh Gyan Vihar University, Jaipur, India}
\author{M. K. Gupta}
\email{mkgupta72@gmail.com} \affiliation{Department of Electrical Engineering, Suresh Gyan Vihar University, Jaipur, India}
\author{Anil Kumar Yadav}
\email{abanilyadav@yahoo.co.in} \affiliation{Department of Physics, United College of Engineering and Research, Greater Noida – 201 306, India}
\author{G. K. Goswami}
\email{gk.goswami09@gmail.com} \affiliation{Department of Mathematics, Netaji Subhas University of Technology, Delhi, India}

%%%%%%%%%%%%%%%%%%%%%%%%%%%%%%%%%%
\begin{abstract}
In this paper, we have proposed a model of accelerating Universe with binary mixture of bulk viscous fluid and dark energy. 
and probed the model parameters: present values of Hubble's constant $H_{0}$, Equation of state paper of dark energy $\omega_{de}$ and density parameter of dark energy $(\Omega_{de})_{0}$ with recent OHD as well as joint Pantheon compilation of SN Ia data and OHD. Using cosmic chronometric technique, we obtain $H_{0} = 69.80 \pm 1.64~km~s^{-1}Mpc^{-1}$ and $70.0258 \pm 1.72~km~s^{-1}Mpc^{-1}$ by restricting our derived model with recent OHD and joint Pantheon compilation SN Ia data and OHD respectively. The age of the Universe in derived model is estimated as $t_{0} = 13.82 \pm 0.33\; Gyrs$. Also, we observe that derived model represents a model of transitioning Universe with transition redshift $z_{t} = 0.7286$. We have constrained the present value of  jerk parameter as $j_{0} = 0.969 \pm 0.0075$ with joint OHD and Pantheon data. From this analysis, we observed that the model of the Universe, presented in this paper shows a marginal departure from $\Lambda$CDM model.
\end{abstract}



\maketitle
\section{Introduction} 
The interval between the end of the $20^{th}$ century and the beginning of the $21^{st}$ century is the golden
time for cosmology. Some remarkable researches in this period $viz$ the observation of late time cosmic acceleration \cite{Riess/1998,Perlmutter/1999}, the measurement of neutrino oscillations, and the detection gravitational waves abruptly changed the traditional concept of our Universe and opened new windows in front of the scientific community. We note that the observational data is the the key ingredient for these discoveries, and the cosmology, we are looking today, has become more enlightening and exact. The major contribution of dark energy (DE) in energy budget of the Universe is the most exciting concept in this period which is truly needed to explain the late time accelerated expansion of the Universe. Therefore, understanding the late time accelerating behavior of the Universe and involved dynamics has remained one of the major challenges in cosmology. In order to address this issue, various cosmological models have been proposed and investigated over the last several years. Among all these cosmological models, the cosmological constant cold dark matter ($\Lambda$CDM) model - one of the most simplest cosmological model which fits excellently to the recent observational probes. However, the involved physics of dark Universe as we as cosmological constant problem, is a serious issue in $\Lambda$CDM model \cite{Weinberg/1989}. Additionally, the recently investigated several anomalies and tensions between different cosmological probes (if such anomalies are not due to the systematic errors) may put straightforward questions on the $\Lambda$CDM model. The dark matter (DM) and DE are assumed to be independent ingredients in $\Lambda$CDM Universe but the physics of dark components of the Universe is not yet clear. Thus there should not be any reason to exclude the possibility of an interaction or energy exchange between DM and DE. In Ref. \cite{Amendola/2000}, the author has investigated cosmological model with interacting phenomenon and generalized the non- interacting scenarios while in Refs. \cite{Campo/2009,Velten/2014}, the authors have given a clue that an interaction between DM and DE may provide a promising solution to the cosmic coincidence problem. Additionally, an interaction between DM and DE can provide a possible solution to $H_{0}$ tension \cite{Kumar/2017,Hassan/2020,Valentino/2017,Yang/2018a,Valentino/2020pdu} and $S_{8}$ tension \cite{Pourtsidou/2016,An/2018} - the $H_{0}$ arises between the CMB measurements by Planck satellite and $S_{8}$ tension arises between the Planck and weak lensing measurements \cite{Valentino/2020IntertwinedIII}. However, in the literature, the possible solutions of the problems associated with $\Lambda$CDM model have been already proposed in the form of holographic dark energy (HDE) models. THe HDE model is based on the holographic principle (HP) and quantum theory of gravity. Li \cite{Li/2004} has investigated that the HDE model with Hubble horizon as infra-red cutoff does not favor the late time acceleration of the Universe but with some other infrared cutoffs e. g. event horizon, particle horizon, etc, the HDE models are able to explain the current accelerated expansion of the Universe\cite{Zhang/2005,Feng/2007,Li/2009a,Luongo/2017,Malekjani/2018,Setare/2006,Setare/2007,Saridakis/2008,Cai/2007,Gao/2009,Wei/2008a}. The general approach to some specific HDE models and its generalization is given in Refs. \cite{Nojiri/2019,Nojiri/2020}. In our recent paper, we have discussed the problems associated with some specific type HDE models in the frame work of modified theories of gravity\cite{Yadav/2021epjc,Yadav/2020arxiv}. In this paper, our aim is to describe the late time acceleration of the Universe beyond $\Lambda$CDM and HDE models. \\ 

In 1940, firstly, Eckart had investigated a bulk viscous cosmological model based on only first order deviation from 
equilibrium \cite{Eckart/1940}. Later on, Israel and Stewart \cite{Israel/1976} have formulated the relativistic second-order theory for bulk viscous fluid. The bulk viscosity appears any time a fluid expands quickly and ceases to be in thermodynamic equilibrium. Thus, the bulk viscosity is appraise of the pressure required to restore equilibrium to a compressed form of expanding system. Some general applications of bulk viscosity in contexts of various cosmological models have been in Refs. \cite{Weinberg/1972,Nightingale/1973,Heller/1975,Heller/1973,Murphy/1973,Burd/1994,Maartens/1995,Zimdahl/1996,Zimdahl/1996a,Chimento/2000,Chimento/2003}. After discovery of accelerating expansion of the Universe at present epoch which is explained with the negative pressure of the cosmic fluid, a bulk viscous pressure $\bar{p}=p-3\xi H^2$, $\xi$ denotes the coefficient of bulk viscosity, is appealing to reconstruct a model of accelerating Universe with inclusion of bulk viscosity. So, here we are intending to propose a model of accelerating Universe with binary mixture of bulk viscous fluid and dark energy. 
It is worthwhile to note here that the viscous fluid cosmological models are instrumental in order to discuss the observed smooth matter distribution on the high entropy per baryon. Recently, Brevik et al. \cite{Brevik/2017,Brevik/2017a} have investigated that the bulk viscosity may leads an inflationary phase of expansion of the early Universe. In Cataldo et al. \cite{Cataldo/2005}, the possibilities of late time acceleration of the Universe has been analyzed. In this paper, the authors have investigated the possible existence big-rip singularity by taking into account an ansatz for the Hubble parameter. Piattella et al. \cite{Piattella/2011} have investigated a bulk viscous Universe aiming to unified DM and DE. The dynamical behavior of bulk viscous matter dominated Universe in Israel–Stewart theory is presented by  Jerin-Mohana et al. \cite{Mohana/2017}. The idea of having the bulk viscosity governs the acceleration of the Universe is given by Padmanabhan and Chitre \cite{Padmanabhan/1987}. The studies of unified dark matter models postulate an exotic equation of state for viscous matter of the form $\xi = \xi_{0}\rho^{n}$ ($\xi_{0}$, n are parameters) \cite{Murphy/1973} and investigate its effect on different expanding phases of the Universe \cite{Fabris/2006,Colistete/2007,Li/2009}. In Ref. \cite{Kamenshchik/2001}, it has been shown that Chaplygin gas model behaves like a special case of the bulk viscous model. The main conclusion of above studies specially in Refs. \cite{Li/2009}, the authors have investigated that both the DM and DE can not explain by using a single viscous fluid. This means that the bulk viscosity becomes important at low redshifts in order to have the negative pressure. Some important applications of bulk viscosity in contexts of general relativity and extended theories of gravity are given in Refs. \cite{Wilson/2007,Mathews/2008,Jeon/1995,Yadav/2019bulk,Bhardwaj/2019,Odintsov/2020,Carlevaro/2005,Singh/2014,Mishra/2018,Ahmed/2019,Yadav/2012}.  \\

The physical cosmology is based on the cosmological principle which states that our Universe is spatially homogeneous and isotropic \cite{Hubble/1929,Hubble/1931}. The modeling of physical cosmology had began on the basic of cosmological principle by Weyl \cite{Weyl/1923}, Friedmann\cite{Friedmann/1922,Friedmann/1924} and Lemaiture\cite{Lemaitre/1927}. The final geometrical shape was given by Robertson\cite{Robertson/1935} and Walker\cite{Walker/1936}. Therefore, the homogeneous and isotropic cosmological model of the Universe was formulated which is commonly known as Friedmann Lemaiture Robertson Walker (FLRW) cosmological model. According to the FLRW models, the Universe had gone in the past through a space-time singularity at a finite time. This singularity is known as big bang singularity and it is an unavoidable event. The entire Universe evolved after big bang and all the laws of physics and mathematics breakdown at this junction. We do not actually know what had happened during and before the big bang singularity. Recent observations indicates that our Universe is nearly flat which is described by FLRW space-time. \\

Inspired by the above researches, in this paper, we propose a model of accelerating Universe other than $\Lambda$CDM model by considering the bulk viscous fluid and DE fluid as the source of matter/energy in the FRW background
expansion. The motivation of constructing this model is due to the quintessence models $(\omega_{de} > -1)$ \cite{Fujii/1982,Carroll/1998} and phantom DE models $(\omega_{de} < -1)$ \cite{Cadwell/2003} of the Universe. Also, in Ref. \cite{Padmanabhan/1987}, it has been investigated that the bulk viscosity might be produce acceleration in the inflationary era of the early Universe. The accelerating phenomenon of late Universe has been investigated by several authors by taking in account the effect of bulk viscosity \cite{Avelino/2010,Sasidharan/2015,Sasidharan/2016,Mohana/2017,Goswami/2021}. In a recent review, Brevik et al. \cite{Brevik/2017} have shown important implications and capabilities of viscosity in describing inflationary era and present epoch of the Universe. The present work is organized as follows. In section 2, we construct the model and its basic equations. Section 3 deals with the method of obtaining observational constraints on model parameters of the viscous Universe. In section 4, the deceleration parameter, age of the Universe and jerk parameter are obtained. The present values of there parameters are also constrained by bounding the derived model with observational data.  Finally in section 5, we have summarized the findings of this paper.        
%%%%%%%%%%%%%%%%%%%%%%%%%%%%%%%%%%%%%%%%%%%%%%%%%%%%%%%%%%%%%%
\section{The model and basic formalism}
We assume the background to be a spatially homogeneous and isotropic FRW space-time
\begin{equation} \label{eq1}
ds^{2}= dt^{2}- a(t)^{2}\left(dx^{2} + dy^{2} + dz^{2}\right).
\end{equation} 
where $a(t)$ is scale factor that describes the rate of expansion of the Universe. Also, in this paper we use units $c = 1$.\\

In this paper, we proposed a model of the Universe that is filled with bulk viscous matter and DE fluid. Therefore, the energy momentum tensor is read as
\begin{equation}\label{E-2}
T_{ij}= T_{ij}^{m} + T_{ij}^{de}.
\end{equation}
where, $T_{ij}^{m}$ and $T_{ij}^{de}$ stand for the energy momentum tensors of bulk viscous matter and DE fluid respectively. Hence,
%%%%%%%%%%%%%%%%%%%%%%%%%%%%%
\begin{eqnarray}
T_{ij}^{m} &=& (\rho+\bar{p})u_iu_j-\bar{p} g_{ij}\nonumber\\
           &=& (\rho + {p-3\xi H^2 }) u_{i}u_{j}- {(p- 3\xi H^2) } g_{ij},\\
T_{ij}^{de} &=& diag[\rho^{de}, -p^{de}, -p^{de}, -p^{de}].
\end{eqnarray}
where $\bar{p}=p-3\xi H^2$ is the pressure of the bulk viscous matter. $\xi$ is the bulk viscous coefficient while $p$ and $H$  denotes the pressure of perfect fluid and Hubble rate respectively. $\rho$ and $\rho_{de}$ are the energy densities of matter field and DE respectively. $p_{de}$ denotes the pressure of DE. Since, the current Universe is filled with pressure-less matter (dust) which has energy density $\rho_{m}$ and its pressure is zero. Hence, $\bar{p}= - 3\xi H^2$ . \\

Therefore, Einstein's field equations for model \eqref{eq1} are obtained as 
\begin{equation}
\label{1}
2\frac{\ddot{a}}{a}+\frac{\dot{a}^{2}}{a^{2}} = -8\pi(p_{de}-3\xi H^2),
\end{equation}
\begin{equation}
\label{2}
3\frac{\dot{a}^{2}}{a^{2}} = 3H^{2} = 8\pi(\rho_m+\rho_{de}).
\end{equation}
From Eq. (\ref{2}), we obtain
\begin{equation}\label{3}
\Omega_{m} + \Omega_{de}=1.
\end{equation}
where $\Omega_m=\frac{8\pi\rho_m}{3H^2}$ and $\Omega_{de}= \frac {8\pi\rho_{de}}{3H^2}$ denote the density parameters of matter and DE fluid respectively.\\
%%%%%%%%%%%%%%%%%%%%%%%%%%%%%
Therefore, the energy conservation equation is read as
\begin{equation} \label{ece}
T_{;j}^{ij}=\dot{\rho_{eff}}+3H(p_{eff}+\rho_{eff})=0.
\end{equation} 
where, $\rho_{eff}=\rho_{m}+\rho_{de}$ and $p_{eff}=p_{de}-3\xi H^2$. We consider the non interacting case of the bulk viscous fluid and DE, therefore, the energy conservation equation $\dot{\rho_{m}}+\dot{\rho_{de}}+3H(\bar{p}+ p_{de} + \rho_{m}+\rho_{de})=0$ for both the fluids are conserved separately.  
\begin{equation} \label{EC-1}
\dot{\rho}_{m}+3H(\rho_{m}-3\xi H^2)=0. 
\end{equation}
\begin{equation} \label{EC-2}
\dot{\rho}_{de}+3H(p_{de}+\rho_{de})=0.
\end{equation}
where $p_{de} = \omega_{de}\rho_{de}$. $\omega^{de}$ is the equation of state parameter of dark energy.\\
The scale factor is in terms of red-shift is given as
\begin{equation}\label{a-z}
a = \frac{a_{0}}{1+z}.
\end{equation}
Here, $a_{0}$ denote the value of scale factor at present epoch.\\
Eqs. (\ref{EC-2}) and (\ref{a-z}) lead to
\begin{equation}
\label{de-1}
\rho_{de} = (\rho_{de})_0(1+z)^{1+\omega_{de}}.
\end{equation}
Using Eqs. (\ref{EC-1}) - (\ref{de-1}), Eq. (\ref{2}) lead to
\begin{equation}\label{H-1}
3(8\pi\xi-1)H^2+(1+z)(H^2)'= 3H_0^2\omega_{de}(\Omega_{de})_0(1+z)^{3(1+\omega_{de})}.
\end{equation}
Thus, the expression for Hubble's function and deceleration parameter are given as
\begin{equation}\label{H-2}
H = H_0 \sqrt { \frac{\omega _{\text{de}} (\Omega_{\text{de}})_0} {8\pi \xi+\omega_{de}}(1+z)^{3(1+\omega_{de})} +  \frac{\omega _{\text{de}} (\Omega _{\text{m}})_0+8\pi \xi}{8\pi \xi+\omega_{de}} (1+z)^{3(1-8\pi\xi)}}. 
\end{equation}
It is worthwhile to note that in absence of bulk viscosity i. e. $\xi = 0$, Eq. (\ref{H-2}) leads to
\begin{equation}
\label{HR-1}
H = H_{0}\sqrt{(\Omega _{\text{m}})_0(1+z)^{3}+(\Omega_{\text{de}})_0(1+z)^{3(1+\omega_{de})}}
\end{equation}
Also, we observe that for $\omega_{de} = -1$, Eq. (\ref{HR-1}) becomes
\begin{equation}
\label{HR_2}
H = H_{0}\sqrt{(\Omega _{\text{m}})_0(1+z)^{3}+(\Omega_{\Lambda})_0}
\end{equation}
where $(\Omega_{\Lambda})_0 = (\Omega_{\text{de}})_0$.\\
Thus, in absence of bulk viscosity $(\xi = 0)$ and $\omega_{de} = -1$. Therefore, the proposed model of the Universe describes$\Lambda$CDM model.\\

The expression for luminosity distance, distance modulus and apparent distance $(m_{b})$ are respectively read as 
\begin{equation}
\label{D-1}
D_{L}=\frac{(1+z)}{H_{0}}\int^z_0\frac{dx} {\sqrt {  \frac{\omega _{\text{de}} (\Omega_{\text{de}})_0} {8\pi \xi+\omega_{de}}(1+x)^{3(1+\omega_{de})} +  \frac{\omega _{\text{de}} (\Omega _{\text{m}})_0+8\pi \xi}{8\pi \xi+\omega_{de}} (1+x)^{3(1-8\pi\xi)}}},
\end{equation}
\begin{equation}\label{D-2}
\mu = 25+ 5log_{10}\left(\frac{(1+z)}{H_{0}}\int^z_0\frac{dx}{\sqrt {  \frac{\omega _{\text{de}} (\Omega_{\text{de}})_0} {8\pi \xi+\omega_{de}}(1+x)^{3(1+\omega_{de})} +  \frac{\omega _{\text{de}} (\Omega _{\text{m}})_0+8\pi \xi}{8\pi \xi+\omega_{de}} (1+x)^{3(1-8\pi\xi)}}} \right)
\end{equation}
\begin{equation}\label{D-3}
m_b = 16.08+ 5log_{10}\left(\frac{1+z}{.026} \int^z_0\frac{dx}{\sqrt { \frac{\omega _{\text{de}} (\Omega_{\text{de}})_0} {8\pi \xi+\omega_{de}}(1+x)^{3(1+\omega_{de})} + \frac{\omega _{\text{de}} (\Omega _{\text{m}})_0+8\pi \xi}{8\pi \xi+\omega_{de}} (1+x)^{3(1-8\pi\xi)}}}\right)
\end{equation}
%%%%%%%%%%%%%%%%%%%%%%%%%%%%%%%%%%%%%
\section{Observational constraints on model parameters of the viscous Universe}
In this section, we explain the observational $H(z)$ data (OHD) and Pantheon compilation SN Ia data and the statistical technique for probing parameters of the viscous Universe .
\begin{itemize} 
\item {\bf OHD}: For the study of cosmic expansion history, the measurement of $H(z)$, due to its model independent nature, becomes useful tool for understanding the recent evolution of the Universe. In this paper, We have considered $46~H(z)$ data points of OHD in the range of $0\leq z\leq 2.36$, obtained from cosmic chronometric technique. These data and its references are given in Table I of Ref. \cite{Biswas/2019}.\\

The $\chi^{2}$ for 46 OHD points is given by
\begin{equation}
\label{chiH}
\chi^{2}_{OHD} = \sum_{i=1}^{46}\left[\frac{H_{th}(z_{i})-H_{obs}(z_{i})}{\sigma_{i}}\right]^{2}
\end{equation}
where $H_{th}(z_{i})$ and $H_{obs}(z_{i})$ denote the theoretical and observed values respectively, and $\sigma_{i}^{2}$ standard deviation of each $H_{obs}(z_{i})$.

\item {\bf Pantheon data}: In this paper, we consider latest Pantheon compilation of SN Ia data \cite{Scolnic/2018} that includes 276 SNIa $(0.03 < z < 0.65)$. 
The  $\chi^{2}$ for Pantheon data is read as
\begin{equation}
\label{chiS}
\chi^{2}_{Pantheon} = m^{T}C^{-1}m,
\end{equation}
where $m = m_{B} - m_{th}$ with
\begin{equation}
\label{chiS-1}
m_{th} = 5log_{10}D_{L}+M,
\end{equation}  
where $C$ is the co-variance matrix of $\mu_{obs}$ \cite{Conley/2011} and the other symbols have their usual meaning.
\end{itemize}
%%%%%%%%%%%%%%%%%%%%%%%%%%
For joint analysis, $\chi^{2}_{joint}$ is computed as
\begin{equation}
\label{joint}
\chi^{2}_{joint} = \chi^{2}_{H(z)} + \chi^{2}_{Pantheon}
\end{equation}
The uniform priors for free parameters are considered as\\
\begin{equation}
\label{priors}
H_{0}\sim U(55,\; 85),\;\; (\Omega_{de})_{0} \sim U(0.55,\; 0.85)\;\;\; \omega_{de} \sim U (-0.6, -1.4) \&\;\; \xi \sim U (-0.03,\; 0.03)
\end{equation}
%%%%%%%%%%%%%%%%%%%%%%%%%%
\begin{figure}[ht!]
\centering
\includegraphics[width=8cm,height=7cm,angle=0]{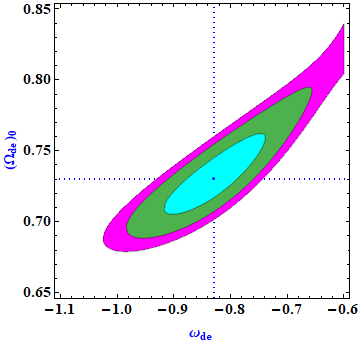} 
\caption{2D contours at $1\sigma$, $2\sigma $ and  $3\sigma$ confidence levels using 46 OHD points. }
\end{figure}
%%%%%%%%%%%%%%%%%%%%%%%%%%%%%%%%%%%%
%%%%%%%%%%%%%%%%%%%%%%%%%%
\begin{figure}[ht!]
\centering
\includegraphics[width=8cm,height=7cm,angle=0]{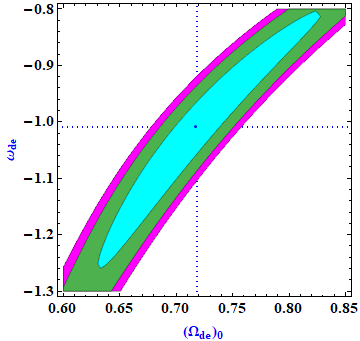}  
\caption{2D contours at $1\sigma$, $2\sigma $ and  $3\sigma$ confidence levels using joint OHD and pantheon compilation of SN Ia data. }
\end{figure}
%%%%%%%%%%%%%%%%%%%%%%%%%%%
%%%%%%%%%%%%%%%%%%%%%%%%%%
\begin{figure}[ht!]
\centering
\includegraphics[width=8cm,height=7cm,angle=0]{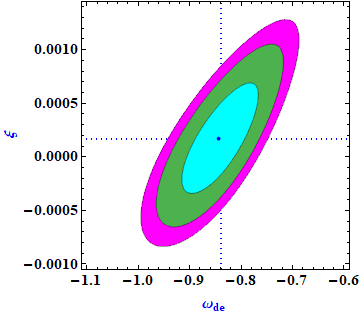} 
\caption{2D contours at $1\sigma$, $2\sigma $ and  $3\sigma$ confidence levels using 46 OHD points. }
\end{figure}
%%%%%%%%%%%%%%%%%%%%%%%%%%%%%%%%%%%%
%%%%%%%%%%%%%%%%%%%%%%%%%%
\begin{figure}[ht!]
\centering
\includegraphics[width=8cm,height=7cm,angle=0]{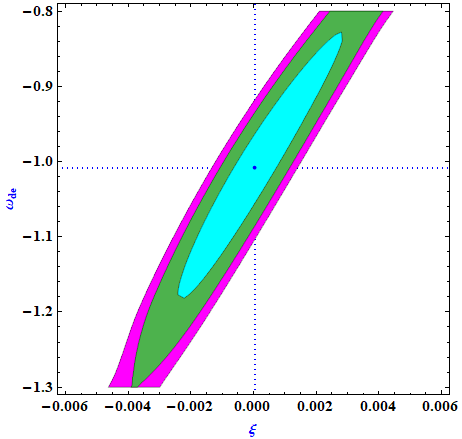}  
\caption{2D contours at $1\sigma$, $2\sigma $ and  $3\sigma$ confidence levels using joint OHD and pantheon compilation of SN Ia data.}
\end{figure}
Figures 1 and 3 represent 2D contours in $\omega_{de}\; - \; (\Omega_{de})_{0}$ plane and $\omega_{de}\;-\;\xi$ plane by analyzing the Universe in derived model with OHD respectively. Also, figures 2 and 4 show 2D contours in $(\Omega_{de})_{0}\;-\; \omega_{de}$ plane and $\xi\; -\; \omega_{de}$ plane by bounding our model with joint OHD and Pantheon compilation SN Ia data respectively. Using OHD data, the best fit values of parameters are obtained as $H_0=69.80 \pm 1.64~km~s^{-1}Mpc^{-1}$, $(\Omega_{de})_{0} = 0.731 \pm 0.013$, $\omega_{de} = -0.83 \pm 0.09$ and $\xi = 0.00016\pm 0.0005$ along with $\chi^{2}_{min} = 23.156$ and $\frac{\chi^{2}_{min}}{N-\eta} = 0.5385$. Here $N$ and $\eta$ stand for total number of data and number of free parameters respectively. It is worthwhile to note that to determine the best model between variety of models, Akaike Information Criterion (AIC) \cite{Akaike/1974} and Bayesian Information Criterion (BIC) \cite{Schwarz/1978} have been proposed. AIC = $\chi^{2} +2\eta$ and BIC = $\chi^{2} + \eta~ln~N$ (see Refs. \cite{Singla/2020mpla,Liddle/2004,lowski/2006,Biesiada/2007}). Thus, we obtain AIC = 6.5385 and BIC = 5.5185 for derived model bounding with OHD data. Similarly for joint Pantheon data and OHD, the best fit values of parameters are obtained as $H_{0} = 70.0258 \pm 1.72;km\;s^{-1}\;Mpc^{-1}$, $(\Omega_{de})_{0} = 0.718 \pm 0.018$, $\omega_{de} = -1.009 \pm 0.16$ and $\xi = 0.000034\pm 0.000035$ along with $\chi^{2}_{min} = 1036.328$, $\frac{\chi^{2}_{min}}{N-\eta} = 0.9917$, AIC = 6.9917 and BIC = 10.0527.
%%%%%%%%%%%%%%%%%%%%%%%%%%%%%%%%%%%%
\section{Physical properties of the viscous universe}
\subsection{Deceleration parameter}
The expression for deceleration parameter is read as
\begin{equation}\label{q-1}
q = -1+\frac{(1+z)H^{\prime}(z)}{H(z)}.
\end{equation}
where $H^{\prime}(z) = \frac{dH(z)}{dz}$.\\
Eqs. (\ref{H-2}) and (\ref{q-1}) lead to
\begin{equation}\label{q-2}
q = \frac{1}{2}\left[1 +\frac {3\omega_{de}(\Omega_{de})_0(1+z)^{3(1+\omega_{de})}}{  \frac{\omega _{\text{de}} (\Omega_{\text{de}})_0} {8\pi \xi+\omega_{de}}(1+z)^{3(1+\omega_{de})} +  \frac{\omega _{\text{de}} (\Omega _{\text{m}})_0+8\pi \xi}{8\pi \xi+\omega_{de}} (1+z)^{3(1-8\pi\xi)}} -24 \pi \xi\right].
\end{equation}
\begin{figure}[ht!]
\centering
\includegraphics[width=10cm,height=8cm,angle=0]{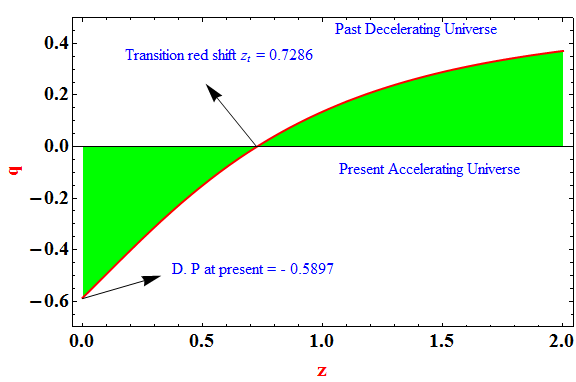} 
\caption{Deceleration parameter $q$ versus red shift $z$.}
\end{figure}
%%%%%%%%%%%%%%%%%%%%%%%%%%%%%%%%%
Fig. 5 depicts the dynamical behavior of deceleration parameter against redshift. We observe that the Universe in derived model was in decelerating phase for $z > 0.7286$. At $z = z_{t} = 0.7286$, the Universe enters into an accelerating phase of expansion and continue to accelerate till now. The present value of deceleration parameter is obtained as $q_{0} =  -0.5897$. It is worthwhile to note that in Capozziello et al. \cite{Capo2020}, the authors have obtained $q_{0} = -0.56 \pm 0.04$ which is quite closer to the value of $q_{0}$, constrained in this paper. Recently, the values of transition redshift are constrained as $z_t=0.74\pm 0.05$ \cite{Farooq2013}, $z_t=0.82\pm 0.08$ \cite{Busca2013}, $z_t=0.69^{+0.23}_{-0.12}$ \cite{Lu2011}, $z_t=0.60^{+0.21}_{-0.12}$ \cite{Yang2019}, $z_t=0.7679^{+0.1831}_{-0.1829}$ \cite{Capo2014} and $z_{t} = 0.72\pm0.04$ \cite{Prasad/2021ijmpa}.   
%%%%%%%%%%%%%%%%%%%%%%%%%%%%%%%%%%%%%%%%%
\subsection{Age of the Universe}
The present age of the Universe is computed as
\begin{equation}\label{23}
H_0 (t_0-t) = \int_{0}^{z}\frac{dx}{(1+x)\left(\sqrt { \frac{\omega _{\text{de}} (\Omega_{\text{de}})_0} {\alpha}(1+x)^{3(1+\omega_{de})} +  \frac{\omega _{\text{de}} (\Omega _{\text{m}})_0+8\pi \xi}{\alpha} (1+x)^{3(1-8\pi\xi)}}\right)}
\end{equation}
As $z \rightarrow \infty$, $t \rightarrow 0$, Eq. (\ref{23}) is recast as following
\begin{equation}\label{age-1}
 t_0 = \lim_{z \rightarrow\infty}{\int_{0}^{z}\frac{dx}{H_0(1+x)\left(\sqrt{\frac{\omega _{\text{de}}(\Omega_{\text{de}})_0}{\alpha}(1+x)^{3(1+\omega_{de})}+\frac{\omega _{\text{de}} (\Omega _{\text{m}})_0 + 8\pi \xi}{\alpha} (1+x)^{3(1-8\pi\xi)}}\right)}}
\end{equation}
%%%%%%%%%%%%%%%%%%%%%%%%%%%%%%
\begin{figure}[ht!]
\centering
\includegraphics[width=10cm,height=8cm,angle=0]{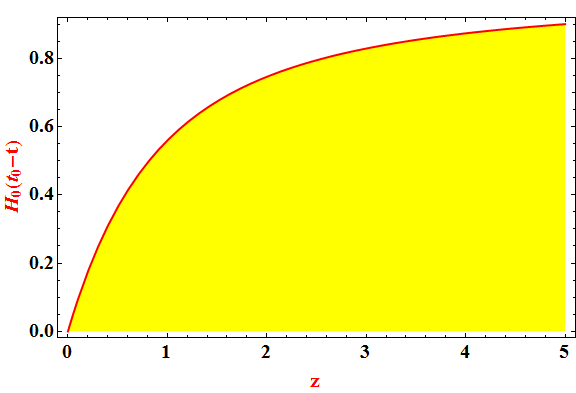} 
\caption{Time $t$ versus red-shift $z$.}
\end{figure}
%%%%%%%%%%%%%%%%%%%%%%%%%%%%%%%%%%%%
Fig. 6 depicts the variation of $H_{0}(t_{0}-t)$ with respect to $z$. The numerical solution of Eq. (\ref{age-1}) leads to $H_{0}t_{0} \sim 0.98732$. Thus, we obtain the present age of the Universe as $t_{0}\sim 0.98732 H_{0}^{-1}$. For joint Pantheon and $H(z)$ data, the present value of Hubble's parameter is estimated as $H_{0} = 70.0258 \pm 1.72;km\;s^{-1}\;Mpc^{-1} \sim 0.06865 \pm 0.0016\; Gyrs^{-1}$. Thus the present age of the Universe in derived model is estimated as $t_{0} = 13.82 \pm 0.33\; Gyrs$. Note that the age of the Universe in Plank results are extracted as $t_0=13.786\pm 0.020$ Gyrs \cite{Planck2020} and $13.81\pm 0.038$ Gyrs \cite{Ade16}. The age of the Universe estimated in this paper is very close to Plank results \cite{Planck2020,Ade16}. Some other estimations for $t_{0}$ are given in Refs. \cite{Goswami/2021,Yadav/2021,Prasad/2020Pramana,Prasad/2020}.    
%%%%%%%%%%%%%%%%%%%%%%%%%%%%%%%%% 
%%%%%%%%%%%%%%%%%%%
\subsection{Jerk parameter} 
The jerk parameter $j=\frac{\dddot{a}}{aH^2}$ describes a suitable kinematic approach to analyze the various cosmological models of the Universe because of its dependence on scale factor. Also, it is useful to investigate a possible deviation from the standard $\Lambda$CDM model. The current values jerk parameter for $\Lambda$CDM model is computed as $j_{0} = 1$.  Any deviation from $j_{0} = 1$ does favor $\Lambda$CDM model \cite{Sahni2003,Alam2003}. The jerk parameters $(j)$ in terms of $q$ is read as
\begin{equation}
j(q)=(2q+1)q+(1+z)\frac{dq}{dz},
\end{equation}
Thus, the jerk parameter in terms of $H$ is expressed as 
\begin{equation}
j(z)=1-(1+z)\frac{\left[H(z)^2\right]^{\prime}}{H(z)^2}+\frac{1}{2}(1+z)^2\frac{ \left[H(z)^2\right]^{\prime\prime}}{H(z)^2}.
\end{equation}
where the prime on the parameters stand for its derivative with respect to $z$. The present value of the jerk parameter is computed as $j_0=(2q_0+1)q_0+\left(\frac{dq}{dz}\right)_{z =0}$, where $q_0$ is the present value of deceleration parameter. In Refs. \cite{Zhai2013,Mamon2018,Mukherjee2017}, the authors have been used different parametrization for jerk parameter and constrained the constants appearing in the parametrization from $H(z)$ data, SN Ia, CMB and BAO data sets. In the present work, we have constrained the present values of jerk parameter from observational $H(z)$ data as well as joint Pantheon and $H(z)$ data without taking any specific parametrization of $j$ in terms of $z$. This is an unbiased approach to constraint the jerk parameter. Figure 7 depicts trajectory of $j$ with respect to $z$ for OHD and joint OHD and Pantheon compilation of SN Ia data.
%%%%%%%%%%%%%%%%%%%%%%%%%%%%%%%%%%%%
\begin{figure}[t!]
\centering
\includegraphics[width=6cm,height=5.5cm,angle=0]{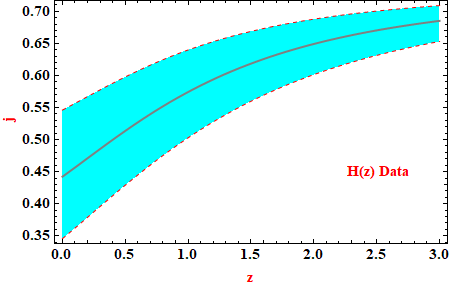}
\includegraphics[width=6cm,height=5.5cm,angle=0]{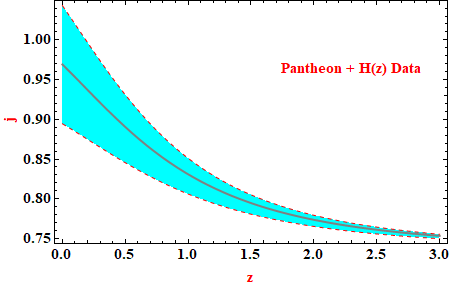} 
\caption{The plot of jerk parameter $j$ versus red-shift $z$.}
\end{figure}
%%%%%%%%%%%%%%%%%%
The value $j_{0}$ extracted from observation $H(z)$ data is $j_{0} = 0.44 \pm 0.013$ which is differ from its value $j_{0} = 1$ for $\Lambda$CDM model. For joint Pantheon and $H(z)$ data, we obtain $j_{0} = 0.969 \pm 0.075$ which is a bit closer to the $\Lambda$CDM value. In Mamon and Bamba \cite{Mamon2018}, jerk parameter is being constrained with observational $H(z)$ data which shows $j_{0}\neq 1$. Some other constraints on $j_{0}$ are given in Refs. \cite{Capo2020,Mukherjee2017,Goswami/2021}. In Ref. \cite{Mukherjee2017}, authors have argued that departure of $j_{0}$ from $1$ favors an interaction between dark component of the Universe.  
%%%%%%%%%%%%%%%%%%%%%%%%%%  
\section{Concluding remarks}
In this paper, we have constructed a model of accelerating Universe with binary mixture of the bulk viscous matter and DE 
fluid and constrained the parameters of the Universe in derived model with recent OHD as well as joint Pantheon compilation of SN Ia data and OHD. We probes the current value of $H_{0}$ as $69.80 \pm 1.64~km~s^{-1}Mpc^{-1}$ and $70.025 \pm 1.72~km~s^{-1}Mpc^{-1}$ by bounding the derived cosmological model with recent $H(z)$ data and joint OHD and Pantheon compilation SN Ia data respectively. Also, we have constrained the present value of jerk parameter $j_{0}$. It is worthwhile to mention that the present value of jerk parameter as as $j_{0} = 0.969 \pm 0.0075$ for bulk viscous accelerating Universe which is differ from its $\Lambda$CDM value $i. e.$ $j_{0}\neq 1$.\\

We conclude here the main features of the derived model. The trajectory of deceleration parameter $q$ with respect to $z$ shows that negative values of $q$ at late times, and positive value of $q$ at an earlier epoch meaning that the Universe in derived model undergoes a transition from a decelerated phase to an accelerated phase of expansion. In our analysis, we obtain transition redshift $z_{t} = 0.7286$. This value of transition redshift is compatible with its values obtained in Refs. \cite{Farooq2013,Busca2013,Lu2011,Yang2019,Capo2014}. It is worthwhile to note that in recent times, there has been significant researches which gives clue that a non - $\Lambda$CDM model may solve $H_{0}$ tension problem \cite{Verde/2019,Poulin/2019,Nunes/2018,Yang/2018,Valentino/2019,Vagnozzi/2019,Vagnozzi/2020,Haridasu/2020,Valentino/2020a,Valentino/2020b}. We have obtained $q_{0} =  -0.5897$ which is quite closer to the recent estimations of $q_{0}$ \cite{Capo2020}. Also we have estimated the present age of the Universe as $t_{0} = 13.82 \pm 0.33$ Gyrs in close vicinity with Plank results \cite{Planck2020,Ade16}.\\

At the end, we conclude that the proposed model of bulk viscous Universe shows a possible departure from the $\Lambda$CDM model. Also. the Universe in derived model is providing reasonable estimates of $t_{0}$, $q_{0}$, $H_{0}$ and $j_{0}$. However, it is important to note that the AIC and BIC criteria in derived model is high which indicates the merit of $\Lambda$CDM model over the proposed model of bulk viscous Universe.
   
%%%%%%%%%%%%%%%%%%%%%%%%%%%%%%%%%%%%%%%%%%%%%%%%%%%%%%%%%%%%%%%%%%%%%%%%%%%%%%%%%%%%%
\section*{Acknowledgments}
The authors are grateful to the anonymous referees for illuminating suggestions that have significantly improved this work in terms of research quality and presentation.

%%%%%%%%%%%%%%%%%%%%%%%%%%%%%%%%%%%%%%%%%

\end{document}